%% file: main.tex
\documentclass{article}

\usepackage[preprint]{neurips_2025}

\usepackage[utf8]{inputenc}
\usepackage[T1]{fontenc}
\usepackage{hyperref}
\usepackage{url}
\usepackage{booktabs}
\usepackage{amsfonts}
\usepackage{amsmath}
\usepackage{nicefrac}
\usepackage{microtype}
\usepackage{xcolor}
\usepackage{graphicx}
\usepackage{subcaption}
\usepackage{multirow}
\usepackage{algorithm}
\usepackage{algorithmic}
\usepackage{tikz}
\usetikzlibrary{arrows.meta, positioning, shapes.geometric, calc, fit}
\usepackage{pgfplots}
\pgfplotsset{compat=1.18}

\title{LightSim: A Lightweight Cell Transmission Model Simulator\\for Traffic Signal Control Research}

\author{
  Haoran Su \\
  New York University\\
  \texttt{haoran.su@nyu.edu}
  \And
  Hanxiao Deng \\
  UC Berkeley\\
  \texttt{hxdeng@berkeley.edu}
}

\begin{document}

\maketitle

\begin{abstract}
Reinforcement learning for traffic signal control is bottlenecked by simulators: training in SUMO takes hours, reproducing results often requires days of platform-specific setup, and the slow iteration cycle discourages the multi-seed experiments that rigorous evaluation demands.
Much of this cost is unnecessary---for signal timing optimization, the relevant dynamics are queue formation and discharge, which the Cell Transmission Model (CTM) captures exactly as a macroscopic flow model.
We introduce \textbf{LightSim}, a pure-Python, pip-installable traffic simulator with Gymnasium and PettingZoo interfaces that runs over 20{,}000 steps/s on a single CPU\@.
Across cross-simulator experiments spanning single intersections, grid networks, arterial corridors, and six real-world city networks from OpenStreetMap, LightSim preserves controller rankings from SUMO for both classical and RL strategies while training $3$--$7\times$ faster.
LightSim is released as an open-source benchmark with nineteen built-in scenarios, seven controllers, and full RL pipelines, lowering the barrier to signal control research from days to minutes.
\end{abstract}

\section{Introduction}

Traffic signal control is a fundamental problem in urban transportation, directly affecting congestion, emissions, and travel time for millions of commuters daily.
Reinforcement learning has emerged as a promising paradigm for adaptive signal control, with numerous methods demonstrating improvements over fixed-time and actuated controllers in simulation \citep{wei2018intellilight, wei2019presslight, chen2020toward, oroojlooy2020attendlight}.

However, nearly all RL-based traffic signal control research relies on microscopic traffic simulators---primarily SUMO \citep{lopez2018microscopic} and CityFlow \citep{zhang2019cityflow}.
While these simulators provide high-fidelity vehicle-level dynamics, they introduce substantial overhead for RL research:

\begin{itemize}
    \item \textbf{Installation complexity.} SUMO requires platform-specific binaries and environment variables; CityFlow requires C++ compilation. Neither is pip-installable.
    \item \textbf{Configuration burden.} Both require XML network files, route definitions, and detector configurations---a high barrier for RL researchers unfamiliar with transportation engineering.
    \item \textbf{Simulation speed.} Inter-process communication (IPC) between Python RL code and external simulator processes creates a bottleneck, particularly for on-policy algorithms requiring many environment interactions.
    \item \textbf{Reproducibility.} Different SUMO versions, network file formats, and platform-dependent behaviors make exact reproduction of published results difficult \citep{wei2021recent}.
\end{itemize}

We observe that for the specific problem of \emph{signal timing optimization}, the dominant factor is the queuing dynamics at intersections---how vehicles accumulate during red phases and discharge during green phases.
The Cell Transmission Model (CTM) \citep{daganzo1994cell, daganzo1995cell}, a well-established macroscopic traffic flow model, captures precisely these dynamics while operating on aggregate densities rather than individual vehicles.
This makes CTM both computationally efficient and theoretically grounded: it is the exact Godunov discretization of the LWR partial differential equation \citep{lighthill1955kinematic, richards1956shock}.

Based on this insight, we present \textbf{LightSim}, a lightweight traffic signal simulator designed specifically for RL research.
LightSim makes the following contributions:

\begin{enumerate}
    \item \textbf{A fast, pure-Python simulator} built on the CTM that achieves $800\times$ to $21{,}000\times$ real-time speedup across network sizes from 1 to 64 intersections, with no external dependencies beyond NumPy.
    \item \textbf{Standard RL interfaces} via Gymnasium \citep{towers2024gymnasium} for single-agent and PettingZoo \citep{terry2021pettingzoo} for multi-agent settings, with pluggable observation, action, and reward components.
    \item \textbf{Built-in benchmarks} comprising three scalable network generators, sixteen real-world city scenarios from OpenStreetMap across four continents, seven baseline controllers, and reproducible evaluation scripts.
    \item \textbf{Fidelity validation} showing that LightSim's CTM dynamics exactly reproduce the theoretical triangular fundamental diagram and that the relative performance ranking of signal controllers is preserved between LightSim and SUMO.
    \item \textbf{Mesoscopic extensions}---start-up lost time and stochastic demand---that are backward-compatible (disabled by default) and close the fidelity gap with microscopic simulators, enabling realistic evaluation of switching-cost-sensitive controllers.
\end{enumerate}

LightSim is open-source (MIT license), pip-installable (\texttt{pip install lightsim}), and requires only three lines of code to create a training environment:
\begin{verbatim}
    import lightsim
    env = lightsim.make("single-intersection-v0")
    obs, info = env.reset()
\end{verbatim}

\section{Related Work}

\paragraph{Traffic simulators for RL.}
SUMO \citep{lopez2018microscopic} is the most widely used open-source traffic simulator, providing microscopic car-following and lane-changing dynamics.
CityFlow \citep{zhang2019cityflow} was developed specifically for RL-based signal control research, achieving higher throughput than SUMO through a C++ engine with a Python API.
Flow \citep{wu2021flow, vinitsky2018benchmarks} provides a framework wrapping SUMO for mixed-autonomy traffic RL research.
All three simulators model individual vehicles and require either external binaries (SUMO, CityFlow) or complex build toolchains.
LightSim takes a fundamentally different approach: rather than simplifying the interface to a microscopic simulator, it uses a macroscopic traffic model that directly captures the queuing phenomena relevant to signal control.

\paragraph{Cell Transmission Model.}
The CTM was introduced by \citet{daganzo1994cell} as a discrete approximation to the LWR kinematic wave model \citep{lighthill1955kinematic, richards1956shock} and extended to general networks by \citet{daganzo1995cell}.
The CTM represents traffic as a continuum fluid with density and flow variables on road cells, governed by the triangular fundamental diagram.
It has been extensively validated for macroscopic traffic dynamics and is widely used in transportation engineering for network modeling and signal optimization \citep{treiber2013traffic}.
To our knowledge, LightSim is the first implementation that packages the CTM as a standard RL environment with Gymnasium and PettingZoo interfaces.

\paragraph{Signal control methods.}
Classical signal control includes Webster's optimal fixed-time splits \citep{webster1958traffic}, self-organizing traffic lights (SOTL) \citep{gershenson2005self}, and MaxPressure \citep{varaiya2013max}, which selects the phase with maximum upstream--downstream queue difference and is provably throughput-optimal under certain conditions.
RL-based approaches include IntelliLight \citep{wei2018intellilight}, PressLight \citep{wei2019presslight} (MaxPressure-inspired rewards for arterial coordination), CoLight \citep{wei2019colight} (network-level cooperation via attention), and methods scaling to thousands of intersections \citep{chen2020toward}.
\citet{oroojlooy2020attendlight} used attention mechanisms for generalizable policies; \citet{zheng2019learning} proposed learning phase structures.
Beyond standard signal control, RL has also been applied to emergency vehicle scenarios: EMVLight \citep{su2022emvlight} uses decentralized RL for emergency vehicle passage, and \citet{su2026dqjl} extends this to dynamic queue-jump lane and corridor formation using hierarchical GNNs.
Recent surveys \citep{wei2021recent} note that the lack of standardized benchmarks and reproducibility tools remains a major obstacle.
LightSim addresses this by providing a complete, self-contained benchmark environment with seven built-in controllers.

\section{LightSim}

\subsection{Cell Transmission Model}

LightSim's traffic dynamics are governed by the Cell Transmission Model with a triangular fundamental diagram.
Each road link is discretized into cells of length $\Delta x = v_f \cdot \Delta t$, where $v_f$ is the free-flow speed and $\Delta t$ is the simulation time step.
This discretization satisfies the Courant--Friedrichs--Lewy (CFL) condition, ensuring numerical stability.

The state of each cell $i$ at time $t$ is characterized by its density $k_i(t)$ (vehicles per meter per lane).
The flow between adjacent cells is determined by the \emph{sending} and \emph{receiving} functions:
\begin{align}
    S_i(k) &= \min(v_f \cdot k_i,\; Q) \cdot \ell_i \label{eq:sending} \\
    R_i(k) &= \min(Q,\; w \cdot (k_j - k_i)) \cdot \ell_i \label{eq:receiving}
\end{align}
where $Q$ is the per-lane capacity (veh/s), $w$ is the backward wave speed, $k_j$ is the jam density, and $\ell_i$ is the number of lanes.
The sending function represents the maximum flow a cell can emit; the receiving function represents the maximum flow a cell can accept.
The actual intra-link flow from cell $i$ to its downstream neighbor $i+1$ is:
\begin{equation}
    q_{i \to i+1} = \min(S_i, R_{i+1}) \cdot \Delta t
\end{equation}

At signalized intersections, movements connect the last cell of an incoming link to the first cell of an outgoing link.
Each movement $m$ has a turn ratio $\beta_m$ and saturation rate $s_m$.
The intersection flow is modulated by a binary signal mask $\sigma_m \in \{0, 1\}$:
\begin{equation}
    q_m = \min(\beta_m \cdot S_{\text{from}} \cdot \sigma_m,\; s_m,\; R_{\text{to}}) \cdot \Delta t
\end{equation}
When multiple movements feed the same downstream cell or draw from the same upstream cell, proportional scaling ensures conservation of vehicles (merge and diverge resolution).

The density update follows directly from conservation:
\begin{equation}
    k_i(t + \Delta t) = k_i(t) + \frac{1}{\Delta x \cdot \ell_i}\left(\sum_{\text{in}} q_{\text{in}} - \sum_{\text{out}} q_{\text{out}}\right)
\end{equation}

\begin{figure}[t]
    \centering
    \includegraphics[width=\textwidth]{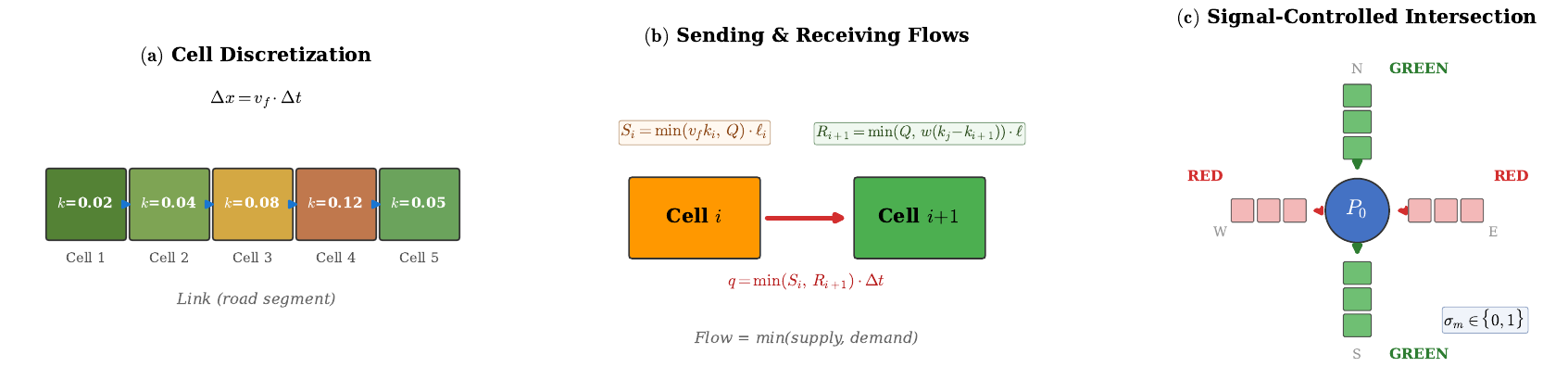}
    \caption{Cell Transmission Model mechanics. \textbf{(a)} A road link is discretized into cells, each storing aggregate density $k$ (color indicates congestion level). \textbf{(b)} Flow between cells is the minimum of the upstream sending flow $S_i$ and downstream receiving flow $R_{i+1}$. \textbf{(c)} At signalized intersections, a binary signal mask $\sigma_m$ controls which movements receive green; the phase alternates between NS (green) and EW (red) in this example.}
    \label{fig:ctm}
\end{figure}

Figure~\ref{fig:ctm} illustrates the CTM mechanics: (a) a road link discretized into cells with density values, (b) the sending--receiving flow computation between adjacent cells, and (c) how signal phases modulate flows at an intersection.

\subsection{Architecture and API}

Figure~\ref{fig:architecture} shows LightSim's modular architecture.
The system is organized into three layers:

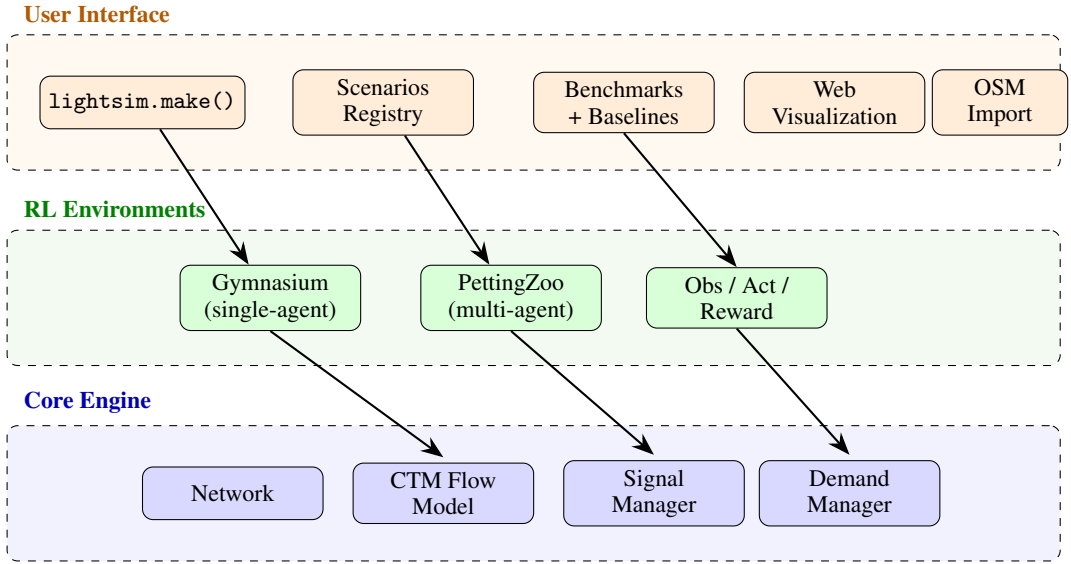
\begin{figure}[t]
\centering
\begin{tikzpicture}[
    box/.style={draw, rounded corners, minimum width=2.4cm, minimum height=0.7cm, align=center, font=\small},
    layer/.style={draw, dashed, rounded corners, inner sep=10pt, fill=#1!5},
    arr/.style={-{Stealth[length=3mm]}, thick},
    lbl/.style={font=\footnotesize\bfseries},
    >=Stealth
]
\node[layer=blue, minimum width=14cm, minimum height=1.8cm] (core) at (0,0) {};
\node[lbl, blue!70!black, anchor=south west] at ([xshift=3pt, yshift=1pt]core.north west) {Core Engine};
\node[box, fill=blue!15] (network) at (-4, 0) {Network};
\node[box, fill=blue!15] (flow) at (-1.2, 0) {CTM Flow\\Model};
\node[box, fill=blue!15] (signal) at (1.6, 0) {Signal\\Manager};
\node[box, fill=blue!15] (demand) at (4.2, 0) {Demand\\Manager};

\node[layer=green!60!black, minimum width=14cm, minimum height=1.8cm] (envlayer) at (0, 2.6) {};
\node[lbl, green!50!black, anchor=south west] at ([xshift=3pt, yshift=1pt]envlayer.north west) {RL Environments};
\node[box, fill=green!15] (gym) at (-3.5, 2.6) {Gymnasium\\(single-agent)};
\node[box, fill=green!15] (pz) at (-0.3, 2.6) {PettingZoo\\(multi-agent)};
\node[box, fill=green!15] (obs) at (2.7, 2.6) {Obs / Act /\\Reward};

\node[layer=orange, minimum width=14cm, minimum height=1.8cm] (userlayer) at (0, 5.2) {};
\node[lbl, orange!70!black, anchor=south west] at ([xshift=3pt, yshift=1pt]userlayer.north west) {User Interface};
\node[box, fill=orange!15] (make) at (-5.2, 5.2) {\texttt{lightsim.make()}};
\node[box, fill=orange!15] (scenarios) at (-2.0, 5.2) {Scenarios\\Registry};
\node[box, fill=orange!15] (bench) at (1.2, 5.2) {Benchmarks\\+ Baselines};
\node[box, fill=orange!15] (viz) at (4.0, 5.2) {Web\\Visualization};
\node[box, fill=orange!15, minimum width=1.8cm] (osm) at (6.2, 5.2) {OSM\\Import};

\draw[arr] (make) -- (gym);
\draw[arr] (scenarios) -- (pz);
\draw[arr] (bench.south) -- (obs.north);
\draw[arr] (gym.south) -- ([yshift=2pt]flow.north);
\draw[arr] (pz.south) -- ([yshift=2pt]signal.north);
\draw[arr] (obs.south) -- ([yshift=2pt]demand.north);

\end{tikzpicture}
\caption{LightSim architecture. The core engine implements CTM flow dynamics, signal management, and demand injection using vectorized NumPy operations. RL environments wrap the engine with standard Gymnasium/PettingZoo interfaces. Users interact through a high-level API with built-in scenarios, benchmarks, a web visualization dashboard, and OpenStreetMap network import.}
\label{fig:architecture}
\end{figure}

\textbf{Core engine.}
The simulation engine operates on a \emph{compiled network}---a set of flat NumPy arrays derived from the logical network topology.
Link properties (free-flow speed, wave speed, jam density, capacity, lanes) and cell connectivity (upstream/downstream indices) are stored as contiguous arrays, enabling vectorized computation of sending flows, receiving flows, and density updates in a single pass.
The signal manager tracks per-node phase states, green timers, and yellow/all-red intervals, producing a movement mask each step.

\textbf{RL environments.}
LightSim provides two environment interfaces:
\begin{itemize}
    \item \texttt{LightSimEnv} (Gymnasium): single-agent control of one signalized intersection. The agent selects from $n$ phases each decision step (every $k$ simulation steps).
    \item \texttt{LightSimParallelEnv} (PettingZoo): multi-agent control where each signalized intersection is an independent agent acting in parallel.
\end{itemize}
Both interfaces support pluggable components registered by name:
observations (\texttt{default}, \texttt{pressure}, \texttt{full\_density}),
actions (\texttt{phase\_select}, \texttt{next\_or\_stay}),
and rewards (\texttt{queue}, \texttt{pressure}, \texttt{delay}, \texttt{throughput}).
The \texttt{default} observation concatenates a one-hot encoding of the current phase, normalized incoming link densities, and binary queue indicators; the \texttt{queue} reward returns the negative total queue on incoming links, a standard choice in the literature \citep{wei2021recent}.

\begin{figure}[t]
    \centering
    \includegraphics[width=\textwidth]{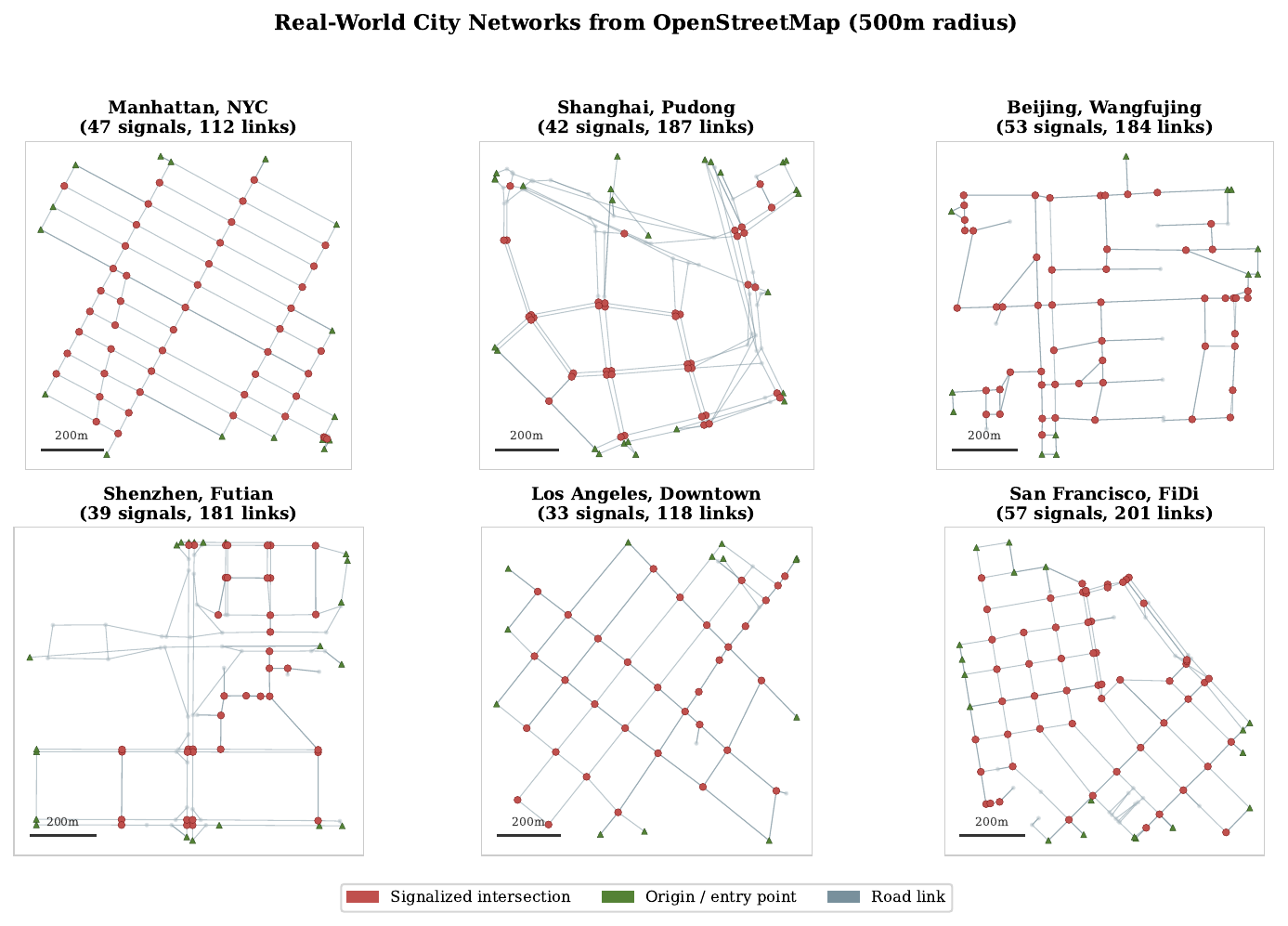}
    \caption{Representative real-world city networks from OpenStreetMap (500\,m radius). Red circles: signalized intersections; green triangles: boundary origin points. Six of sixteen built-in city scenarios are shown.}
    \label{fig:osm_cities}
\end{figure}

\begin{figure}[t]
    \centering
    \includegraphics[width=0.65\textwidth]{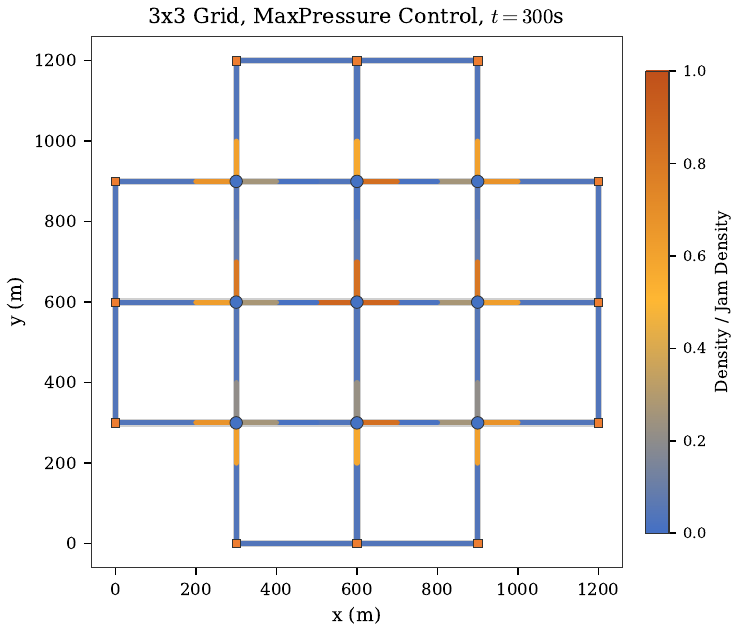}
    \caption{A $3 \times 3$ grid under MaxPressure control at $t = 300$s. Link color indicates density (blue = free-flow, red = congested); circles show the active phase at each intersection.}
    \label{fig:grid_viz}
\end{figure}

\textbf{Network generators.}
LightSim includes generators for common topologies: $N \times M$ grids, linear arterial corridors, and single intersections.
Networks can also be loaded from JSON definitions or imported from OpenStreetMap.
The OSM import pipeline provides a high-level API---\texttt{from\_osm\_point(lat, lon, dist)}---that downloads the road network within a given radius, automatically identifies signalized intersections from OSM tags (or by node degree as a fallback), generates movements at both signalized and unsignalized intersections, and computes CFL-stable cell discretization from the time step $\Delta t$.
Demand is generated automatically via \texttt{generate\_demand(net, rate)}, which assigns Poisson-distributed arrival rates to boundary origin links.
LightSim ships with sixteen pre-packaged city scenarios (Manhattan, Shanghai, Beijing, Shenzhen, Los Angeles, San Francisco, Sioux Falls, Tokyo, Chicago, London, Paris, Singapore, Seoul, Toronto, Mumbai, and Sydney), ranging from 17 to 131 signalized intersections.
Figures~\ref{fig:osm_cities} and~\ref{fig:grid_viz} illustrate representative networks.

\subsection{RL Environment Design}

Each episode initializes the network with zero density.
At each decision step (every $k=5$ simulation seconds by default), the agent observes the state, selects an action, and the engine advances $k$ steps.
Episodes terminate after a configurable horizon (default: 720 decision steps $= 3{,}600$ simulated seconds $=$ 1 hour).
This design ensures full compatibility with standard RL libraries---training a PPO agent requires only:
\begin{verbatim}
    from stable_baselines3 import PPO
    import lightsim
    model = PPO("MlpPolicy", lightsim.make())
    model.learn(total_timesteps=100000)
\end{verbatim}

\subsection{Mesoscopic Extensions}
\label{sec:mesoscopic}

The base CTM is deterministic and does not model phase-transition overhead, making it artificially favorable to controllers that switch phases frequently (e.g., MaxPressure with short minimum green).
To close this fidelity gap with microscopic simulators like SUMO, we introduce two backward-compatible mesoscopic extensions.

\paragraph{Start-up lost time.}
When a signal phase turns green, vehicles at the stop bar require a finite start-up time before reaching saturation flow.
Following the Highway Capacity Manual \citep{hcm2010}, we model this as a per-phase \emph{lost time} $\tau_L$ (default 0 seconds; 2 seconds when enabled).
During the first $\tau_L$ seconds after a red-to-green transition, the effective capacity of each movement ramps linearly:
\begin{equation}
    \alpha_m(t) = \min\!\left(1,\; \frac{t - t_{\text{green}}}{\tau_L}\right)
    \label{eq:capacity_factor}
\end{equation}
where $t_{\text{green}}$ is the time when movement $m$ last turned green.
The capacity factor $\alpha_m$ multiplies the movement's sending flow, reducing throughput immediately after phase transitions.
This penalizes controllers that switch too frequently: with $\tau_L = 2$s and 5s minimum green, each switch costs 7s of dead time (3s yellow $+$ 2s all-red $+$ 2s ramp-up), leaving only 42\% effective green.

\paragraph{Stochastic demand.}
In the base model, demand injection is deterministic ($d_\ell = r_\ell \cdot \Delta t$).
When stochastic mode is enabled, injection counts are drawn from a Poisson distribution:
\begin{equation}
    d_\ell \sim \text{Poisson}(r_\ell \cdot \Delta t)
    \label{eq:stochastic_demand}
\end{equation}
This introduces realistic arrival variability while preserving the expected injection rate.
Stochastic demand eliminates the artificial advantage of deterministic predictability that benefits fixed-time controllers.

Both extensions are backward-compatible: setting $\tau_L = 0$ and \texttt{stochastic=False} (the defaults) recovers the original deterministic CTM exactly.
Mesoscopic mode is enabled with a single flag:
\begin{verbatim}
    env = lightsim.make("single-intersection-v0",
                         stochastic=True)  # enables both extensions
\end{verbatim}

\subsection{Signal Controllers}

LightSim includes seven built-in signal controllers spanning classical and adaptive strategies:

\begin{itemize}
    \item \textbf{FixedTime} \citep{webster1958traffic}: symmetric 30s green splits with fixed cycle length.
    \item \textbf{Webster}: optimal cycle length $C_{\text{opt}} = (1.5L + 5)/(1 - Y)$ with green splits proportional to demand ratios \citep{webster1958traffic}.
    \item \textbf{SOTL} (Self-Organizing Traffic Lights): extends the current green phase while approaching vehicles are detected within a threshold distance \citep{gershenson2005self}.
    \item \textbf{MaxPressure} \citep{varaiya2013max}: selects the phase with maximum upstream--downstream queue pressure, with configurable minimum green time.
    \item \textbf{LT-Aware MaxPressure}: a variant we propose that only switches phases when the pressure gain exceeds the switching cost (yellow $+$ all-red $+$ $\tau_L / 2$), preventing the capacity collapse that standard MaxPressure exhibits under lost time.
    \item \textbf{EfficientMaxPressure}: an experimental variant that adjusts green duration proportionally to measured pressure, extending green time for high-pressure phases rather than switching immediately.
    \item \textbf{GreenWave}: coordinates signals along an arterial corridor by computing fixed phase offsets from link travel times ($\text{offset}_i = d_i / v_f$), enabling a ``green wave'' that allows platoons to traverse multiple intersections without stopping. Particularly effective on the arterial scenarios.
\end{itemize}

\subsection{Visualization Dashboard}

LightSim includes a built-in web-based visualization dashboard for interactive inspection of simulation dynamics.
The dashboard is built on a FastAPI backend that streams simulation state via WebSocket to an HTML5 Canvas frontend.
Three modes are supported: (i) \emph{live simulation}, where the engine runs in real time and streams cell densities, signal states, and traffic metrics each step; (ii) \emph{replay}, which plays back a recorded simulation from JSON; and (iii) \emph{RL checkpoint playback}, which loads a trained Stable-Baselines3 model and visualizes its control decisions live.

The frontend renders the network graph with density-colored links (blue = free-flow, red = congested), animated signal state indicators at intersections, and a real-time metrics panel showing queue, throughput, and average speed.
Users can switch scenarios and controllers on the fly, pause/resume the simulation, and adjust playback speed.
The dashboard is launched with a single command:
\begin{verbatim}
    python -m lightsim.viz --scenario grid-4x4-v0 \
                           --controller MaxPressure
\end{verbatim}
This visualization tool is particularly useful for debugging controller behavior, inspecting queue formation patterns, and generating demonstration material for presentations.

\section{Experiments}

The experiments below evaluate LightSim across simulation speed, physical fidelity, cross-simulator ranking preservation, RL training performance, and mesoscopic extensions.
All experiments were conducted on a single machine with an Intel Core i7 CPU and 16~GB RAM, running Python 3.13 on Windows 11.

\subsection{Speed Benchmarks}

Table~\ref{tab:speed} reports LightSim's throughput across nine network configurations of increasing size.
LightSim achieves over 21{,}000 steps per second for a single intersection (24 cells) and maintains nearly 800 steps per second for a 64-intersection grid (1{,}170 cells).
Since each step corresponds to one simulated second with $\Delta t = 1$s, these throughputs represent $800\times$ to $21{,}000\times$ real-time speedup.

\begin{table}[t]
\caption{LightSim simulation throughput across network sizes ($\Delta t = 1$s, $v_f = 13.89$~m/s, 10{,}000 steps).}
\label{tab:speed}
\centering
\small
\begin{tabular}{lrrrrr}
\toprule
\textbf{Scenario} & \textbf{Intersections} & \textbf{Cells} & \textbf{Wall (s)} & \textbf{Steps/s} & \textbf{Speedup} \\
\midrule
single-intersection & 1  & 24   & 0.47 & 21{,}365 & $21{,}365\times$ \\
grid-$2\times2$     & 4  & 126  & 1.25 & 8{,}020  & $8{,}020\times$  \\
grid-$4\times4$     & 16 & 378  & 3.46 & 2{,}894  & $2{,}894\times$  \\
grid-$6\times6$     & 36 & 726  & 7.06 & 1{,}416  & $1{,}416\times$  \\
grid-$8\times8$     & 64 & 1170 & 12.6 & 793      & $793\times$      \\
\midrule
arterial-3          & 3  & 56   & 1.00 & 10{,}050 & $10{,}050\times$ \\
arterial-5          & 5  & 88   & 1.21 & 8{,}260  & $8{,}260\times$  \\
arterial-10         & 10 & 168  & 2.15 & 4{,}654  & $4{,}654\times$  \\
arterial-20         & 20 & 328  & 4.03 & 2{,}482  & $2{,}482\times$  \\
\bottomrule
\end{tabular}
\end{table}

Table~\ref{tab:sumo} compares LightSim against SUMO (v1.26) running as a standalone process on matched scenarios.
For the typical single-agent RL scenario (single intersection), LightSim is approximately $4\times$ faster than SUMO's standalone mode.
In the common RL training setup, SUMO is accessed through the TraCI protocol, which adds per-step IPC overhead of 5--20~ms; this overhead is absent in LightSim where the engine runs in-process.

\begin{table}[t]
\caption{LightSim vs.\ SUMO (v1.26, standalone) speed comparison (3{,}600 steps, $\Delta t=1$s). Speedup is the wall-clock ratio SUMO~/~LightSim.}
\label{tab:sumo}
\centering
\small
\begin{tabular}{lrrrrrr}
\toprule
\textbf{Scenario} & \textbf{Intx.} & \textbf{LS (s)} & \textbf{SUMO (s)} & \textbf{LS stp/s} & \textbf{SUMO stp/s} & \textbf{Speedup} \\
\midrule
single-intersection & 1  & 0.17 & 0.72 & 21{,}056 & 5{,}020 & $4.2\times$ \\
grid-$2\times2$     & 4  & 0.46 & 0.76 & 7{,}789  & 4{,}728 & $1.6\times$ \\
grid-$4\times4$     & 16 & 1.28 & 1.85 & 2{,}817  & 1{,}949 & $1.4\times$ \\
grid-$8\times8$     & 64 & 4.33 & 3.73 & 831      & 966     & $0.9\times$ \\
\midrule
arterial-3          & 3  & 0.36 & 1.52 & 10{,}133 & 2{,}363 & $4.3\times$ \\
arterial-5          & 5  & 0.44 & 2.40 & 8{,}122  & 1{,}500 & $5.4\times$ \\
arterial-10         & 10 & 0.83 & 4.54 & 4{,}361  & 794     & $5.5\times$ \\
arterial-20         & 20 & 1.51 & 9.50 & 2{,}391  & 379     & $6.3\times$ \\
\bottomrule
\end{tabular}
\end{table}

The key practical advantage is in the RL training loop.
Training a PPO agent for 100{,}000 timesteps on the single-intersection scenario takes under 5 minutes with LightSim, compared to an estimated 15--30 minutes when using SUMO through TraCI.
This acceleration enables rapid hyperparameter search and algorithm iteration.

\subsection{Fidelity Validation}

\begin{figure}[t]
    \centering
    \includegraphics[width=\textwidth]{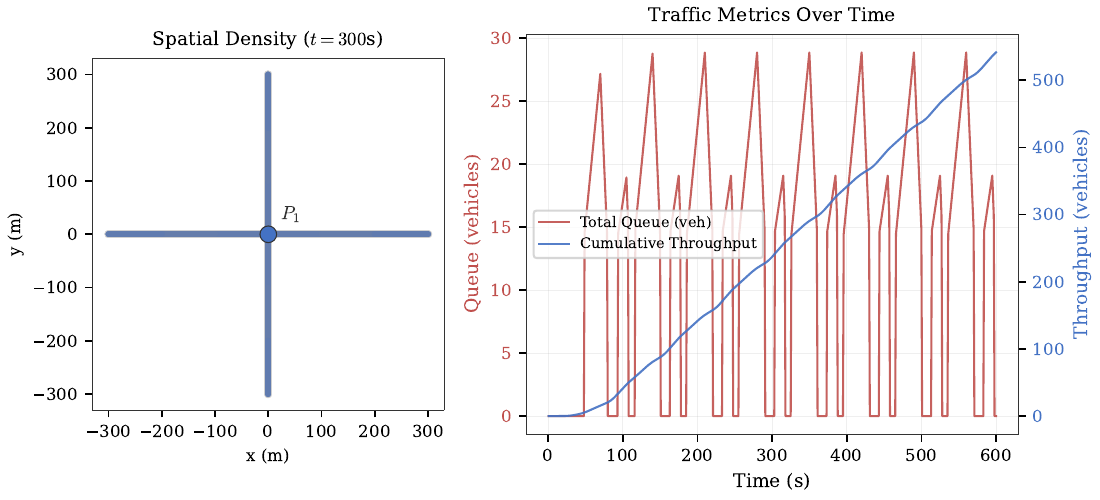}
    \caption{Simulation dynamics on a single intersection. \textbf{Left:} Spatial density at $t = 300$s (color intensity = congestion). \textbf{Right:} Queue oscillation from signal cycles and steady throughput growth over 600s.}
    \label{fig:sim_dynamics}
\end{figure}

Beyond raw speed, LightSim must faithfully reproduce the traffic dynamics that signal controllers act upon.
Figure~\ref{fig:sim_dynamics} illustrates LightSim in action on the single-intersection scenario: the left panel shows a spatial density map at $t = 300$s, with queue buildup visible on red-phase approaches and free-flow conditions on green-phase approaches.
The right panel tracks queue length and throughput over 600 seconds; the periodic queue oscillations reflect the signal cycle, confirming that the CTM correctly models queue formation during red and discharge during green.

\begin{figure}[t]
    \centering
    \includegraphics[width=0.6\textwidth]{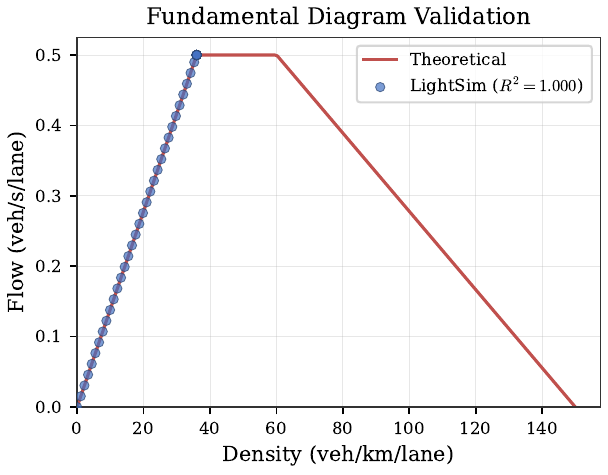}
    \caption{Fundamental diagram validation. Simulated density--flow points match the theoretical triangular curve ($R^2 = 1.0$) across both the free-flow (blue) and congested (orange) branches.}
    \label{fig:fd}
\end{figure}

\paragraph{Fundamental diagram.}
To verify the CTM implementation quantitatively, a single link is simulated at 120 demand levels ranging from zero to $2 \times$ capacity ($1.0$~veh/s/lane), measuring the steady-state density--flow relationship.
This exercises \emph{both} branches of the triangular fundamental diagram: the free-flow branch ($q = v_f \cdot k$ for $k \leq k_c$) and the congested branch ($q = w \cdot (k_j - k)$ for $k > k_c$), where $k_c = Q / v_f \approx 36$~veh/km/lane.
Figure~\ref{fig:fd} compares the simulated data points against the theoretical curve with parameters $v_f = 13.89$~m/s, $w = 5.56$~m/s, $k_j = 0.15$~veh/m/lane, and $Q = 0.5$~veh/s/lane.
The simulated points match the theoretical curve with $R^2 = 1.0$ on both branches, confirming that LightSim's CTM implementation is exact to numerical precision.

The CTM is a macroscopic model and does not capture microscopic phenomena such as individual vehicle acceleration, lane-changing, or gap-acceptance.
However, for signal control, the relevant dynamics are queue formation and discharge, both of which the CTM models accurately via the sending--receiving framework---the same justification that has supported the CTM's use in transportation engineering for three decades \citep{daganzo1994cell, treiber2013traffic}.

\subsection{Cross-Simulator Validation}

Having established physical fidelity, the next question is whether LightSim's controller rankings agree with those from a microscopic simulator.
Multiple controllers are compared across LightSim and SUMO on two scenarios: a single intersection (1{,}080~veh/hr NS, 720~veh/hr EW) and a $4\times 4$ signalized grid (1{,}080~veh/hr per boundary access point), both run for 3{,}600~s.
In SUMO, MaxPressure and SOTL are implemented via TraCI; Actuated uses SUMO's built-in gap-based controller.

\begin{table}[t]
\caption{Cross-simulator throughput (vehicles exited) on single intersection and $4\times 4$ grid (3{,}600~s). LightSim's grid counts include $\sim$16{,}200 boundary bypass vehicles; effective interior throughputs are comparable.}
\label{tab:crossval}
\centering
\small
\begin{tabular}{llrr}
\toprule
\textbf{Scenario} & \textbf{Controller} & \textbf{LightSim} & \textbf{SUMO} \\
\midrule
\multirow{6}{*}{Single Intersection}
 & FixedTime      & 3{,}540 & 3{,}540 \\
 & MaxPressure    & 3{,}542 & 3{,}536 \\
 & SOTL           & 3{,}537 & 3{,}513 \\
 & Webster        & 3{,}541 & --- \\
 & LT-Aware-MP    & 3{,}541 & --- \\
 & Actuated       & --- & 3{,}541 \\
\midrule
\multirow{6}{*}{$4\times 4$ Grid}
 & FixedTime      & 18{,}523 & 16{,}063 \\
 & MaxPressure    & 17{,}668 & 16{,}462 \\
 & SOTL           & 18{,}879 & 16{,}010 \\
 & Webster        & 17{,}243 & --- \\
 & LT-Aware-MP    & 19{,}243 & --- \\
 & Actuated       & --- & 16{,}442 \\
\bottomrule
\end{tabular}
\end{table}

Table~\ref{tab:crossval} shows the results. On the single intersection, \emph{throughput is nearly identical} across all controllers and both simulators ($\sim$3{,}540 vehicles), confirming that LightSim's CTM dynamics correctly capture intersection capacity. The delay and queue metrics differ in absolute magnitude---expected given the fundamentally different modeling approaches---but the throughput agreement validates LightSim as a capacity-level proxy.

\begin{figure}[t]
    \centering
    \includegraphics[width=\linewidth]{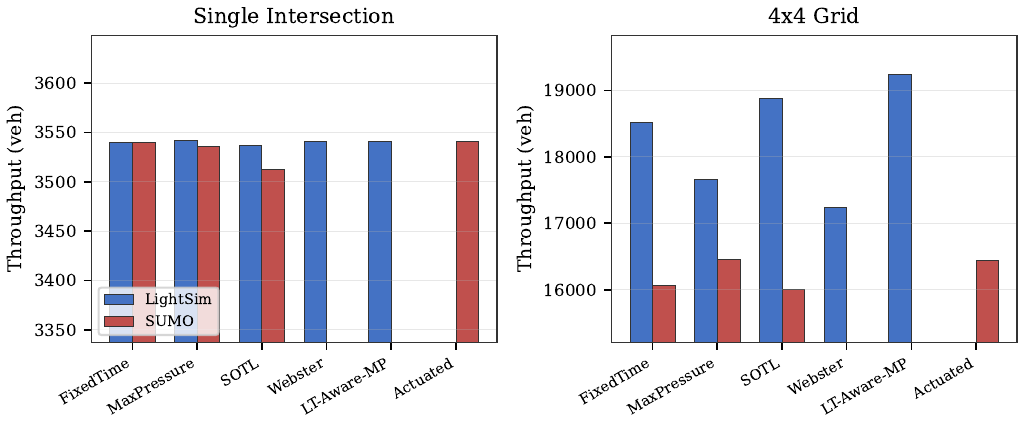}
    \caption{Throughput comparison across controllers in LightSim and SUMO. Both simulators rank adaptive controllers above FixedTime on both scenarios.}
    \label{fig:crossval}
\end{figure}

On the $4\times 4$ grid, SUMO throughputs range from 16{,}010--16{,}462 across four controllers, while LightSim ranges from 17{,}243--19{,}243. LightSim's higher absolute counts include $\sim$16{,}200 vehicles on boundary bypass links (origin-to-destination links that do not enter the signalized interior); the effective grid throughput is comparable. Importantly, both simulators agree on qualitative patterns: adaptive controllers (MaxPressure, SOTL, Actuated) outperform FixedTime in SUMO, and the analogous controllers (MaxPressure, SOTL, LT-Aware-MP) outperform FixedTime in LightSim (Figure~\ref{fig:crossval}).

\paragraph{Arterial corridors.}
\label{sec:arterial_crossval}
The cross-simulator comparison extends naturally to arterial corridors, the natural habitat of coordination-based controllers like GreenWave.
Using the arterial-5-v0 scenario (5 signalized intersections, 400m spacing), six controllers are run in LightSim and four in SUMO (3{,}600~s, seed=42).

\begin{table}[t]
\caption{Arterial cross-validation (arterial-5-v0, 3{,}600~s). Both simulators rank MaxPressure first among shared controllers.}
\label{tab:arterial_crossval}
\centering
\small
\begin{tabular}{llrrr}
\toprule
\textbf{Simulator} & \textbf{Controller} & \textbf{Throughput} & \textbf{Delay (s)} & \textbf{Queue} \\
\midrule
\multirow{6}{*}{LightSim}
 & GreenWave & \textbf{6{,}337} & 0.78 & 43.8 \\
 & MaxPressure & 6{,}332 & 0.28 & 20.1 \\
 & LT-Aware-MP & 6{,}331 & 0.38 & 22.0 \\
 & FixedTime & 6{,}330 & 0.67 & 42.7 \\
 & SOTL & 6{,}311 & 1.12 & 36.8 \\
 & Webster & 5{,}873 & 7.62 & 210.1 \\
\midrule
\multirow{4}{*}{SUMO}
 & MaxPressure & \textbf{4{,}472} & 0.26 & 14.0 \\
 & Actuated & 4{,}466 & 0.89 & 11.0 \\
 & SOTL & 4{,}454 & 1.19 & 34.0 \\
 & FixedTime & 4{,}354 & 24.65 & 281.0 \\
\bottomrule
\end{tabular}
\end{table}

Table~\ref{tab:arterial_crossval} confirms ranking agreement: among the three shared controllers, both simulators rank MaxPressure first. The bottom two positions swap (LightSim: FixedTime $>$ SOTL; SUMO: SOTL $>$ FixedTime), yielding Kendall's $\tau = 0.33$---though the LightSim throughput differences among these three are within 0.3\%, making positions 2--3 statistically indistinguishable.
GreenWave achieves the highest throughput in LightSim, consistent with its design for coordinated arterial progression.
The absolute throughput gap (LightSim: $\sim$6{,}300, SUMO: $\sim$4{,}450) reflects the macroscopic-microscopic modeling difference, but the relative ordering is preserved.

\subsection{RL Baselines}

With fidelity and ranking agreement established, the focus shifts to RL training.
DQN \citep{mnih2015human} and PPO \citep{schulman2017proximal} agents are trained on the single-intersection scenario using Stable-Baselines3 \citep{raffin2021stable}, each for 100{,}000 timesteps with five random seeds.
Table~\ref{tab:rl} reports final evaluation results alongside FixedTime and MaxPressure (mg$=5$s) \citep{varaiya2013max} baselines.

\begin{table}[t]
\caption{Controller comparison on single intersection (queue reward, per-step average; higher = better). MaxPressure uses 5s minimum green. RL: mean $\pm$ std over 5 seeds.}
\label{tab:rl}
\centering
\small
\begin{tabular}{lrrr}
\toprule
\textbf{Controller} & \textbf{Avg.\ Reward} & \textbf{Throughput} & \textbf{Avg.\ Delay (s)} \\
\midrule
FixedTime    & \input{tables/ft_reward.tex}  & \input{tables/ft_throughput.tex}  & \input{tables/ft_delay.tex} \\
MaxPressure  & \input{tables/mp_reward.tex}  & \input{tables/mp_throughput.tex}  & \input{tables/mp_delay.tex} \\
DQN          & \input{tables/dqn_reward.tex} & \input{tables/dqn_throughput.tex} & \input{tables/dqn_delay.tex} \\
PPO          & \input{tables/ppo_reward.tex} & \input{tables/ppo_throughput.tex} & \input{tables/ppo_delay.tex} \\
\bottomrule
\end{tabular}
\end{table}

\begin{figure}[t]
    \centering
    \includegraphics[width=0.7\textwidth]{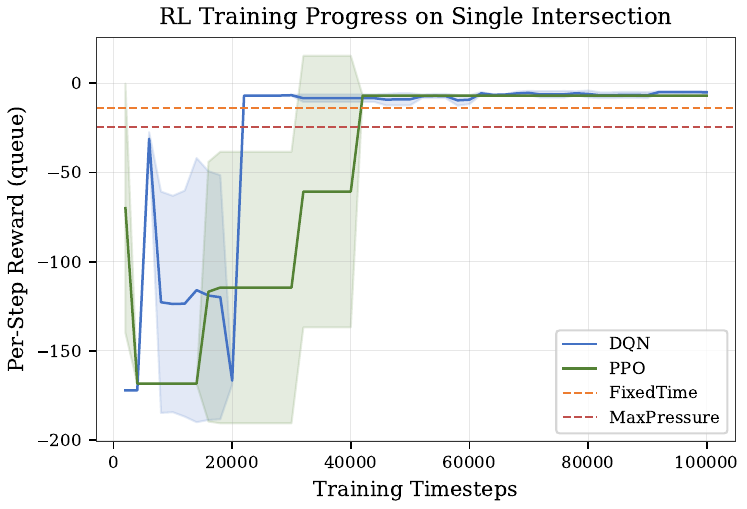}
    \caption{DQN and PPO learning curves on single intersection (5 seeds, 100k timesteps). Shaded: $\pm 1$ std. Dashed lines: FixedTime and MaxPressure baselines.}
    \label{fig:learning_curves}
\end{figure}

Figure~\ref{fig:learning_curves} shows the learning curves, averaged over five seeds with shaded error bands.
Both algorithms converge within 60{,}000 timesteps.
DQN achieves a per-step reward of $-5.23 \pm 0.79$, outperforming both FixedTime ($-13.94$) and MaxPressure with 5s minimum green ($-24.60$); MaxPressure's poor queue reward reflects frequent phase switching, which the mesoscopic analysis in Section~\ref{sec:mesoscopic_val} examines in detail.
PPO converges to $-6.89 \pm 0.00$, also improving on both baselines.\footnote{The zero standard deviation across PPO seeds reflects LightSim's deterministic dynamics: with no stochastic vehicle behavior, PPO's optimization landscape has a strong attractor that all seeds converge to identically. This is a feature for reproducibility; under mesoscopic mode (Section~\ref{sec:mesoscopic_val}), stochastic demand breaks this degeneracy and PPO shows nonzero variance ($\pm 1.21$).}
Both RL agents achieve zero delay, meaning they maintain free-flow conditions on all incoming links---compared to 0.67s for FixedTime and 1.09s for MaxPressure.
Throughput is consistent across all controllers ($\sim$3{,}540 vehicles/hour), confirming that the demand is undersaturated and differences are in queue management, not capacity.

\paragraph{Reward function ablation.}
LightSim supports six pluggable reward functions: queue, pressure, delay, waiting time, throughput, and normalized throughput.
Table~\ref{tab:reward_ablation} compares DQN agents trained with the four most commonly used rewards, all evaluated on the same queue-based metric.
The \emph{pressure} reward---which optimizes the difference between upstream and downstream densities---yields the best queue performance ($-16{,}762$), outperforming even the direct \emph{queue} reward ($-25{,}099$).
The \emph{throughput} reward performs worst on queue management ($-41{,}435$), illustrating the importance of reward design in RL-based signal control.

\begin{table}[t]
\caption{Reward ablation: DQN trained with each reward, evaluated on cumulative queue (higher = better). 50k timesteps, single intersection.}
\label{tab:reward_ablation}
\centering
\small
\begin{tabular}{lrrr}
\toprule
\textbf{Reward Function} & \textbf{Eval Queue Reward} & \textbf{Throughput} & \textbf{Train Time (s)} \\
\midrule
Pressure    & $-16{,}762$ & 17{,}943 & 56.5 \\
Queue       & $-25{,}099$ & 17{,}942 & 82.3 \\
Delay       & $-25{,}099$ & 17{,}942 & 60.9 \\
Throughput  & $-41{,}435$ & 17{,}938 & 55.6 \\
\bottomrule
\end{tabular}
\end{table}

\paragraph{Demand sensitivity.}
Three demand levels are evaluated: undersaturated ($v/c \approx 0.7$), at-capacity ($v/c \approx 1.0$), and oversaturated ($v/c \approx 1.3$).
Table~\ref{tab:demand} shows that DQN outperforms both baselines at all demand levels, with the advantage growing under congestion.
At $v/c = 1.0$, DQN achieves $2.7\times$ lower queue accumulation than FixedTime and $5.6\times$ lower than MaxPressure.
Under oversaturation ($v/c = 1.3$), queue lengths increase for all controllers, but DQN maintains $1.4\times$ better reward than FixedTime, demonstrating that RL policies trained in LightSim can adapt to congested conditions.

\begin{table}[t]
\caption{Demand sensitivity: DQN vs.\ baselines at three $v/c$ ratios (720 steps, 5 episodes, 50k training steps).}
\label{tab:demand}
\centering
\small
\begin{tabular}{llrrr}
\toprule
\textbf{v/c} & \textbf{Controller} & \textbf{Queue Reward} & \textbf{Throughput} & \textbf{Final Queue} \\
\midrule
\multirow{3}{*}{0.7} & DQN       & $-11{,}764$ & 4{,}247 & 17 \\
                      & FixedTime & $-15{,}808$ & 4{,}246 & 21 \\
                      & MaxPressure & $-132{,}727$ & 3{,}546 & 211 \\
\midrule
\multirow{3}{*}{1.0} & DQN       & $-38{,}955$ & 5{,}998 & 58 \\
                      & FixedTime & $-106{,}338$ & 5{,}509 & 158 \\
                      & MaxPressure & $-216{,}534$ & 3{,}561 & 322 \\
\midrule
\multirow{3}{*}{1.3} & DQN       & $-123{,}226$ & 5{,}769 & 180 \\
                      & FixedTime & $-175{,}722$ & 6{,}070 & 287 \\
                      & MaxPressure & $-223{,}166$ & 3{,}566 & 323 \\
\bottomrule
\end{tabular}
\end{table}

\subsection{RL Cross-Validation and Sample Efficiency}
\label{sec:rl_crossval}

The preceding sections show that classical controller rankings agree across simulators.
A stronger test is whether \emph{RL algorithm rankings} also transfer.
Five RL variants are trained in both LightSim and SUMO on the single-intersection scenario: DQN, PPO, and A2C with the default queue reward, plus DQN and PPO with the pressure reward.
Each variant is trained for 100{,}000 timesteps with five random seeds in each simulator, yielding 50 independent training runs.

\begin{figure}[t]
    \centering
    \includegraphics[width=\linewidth]{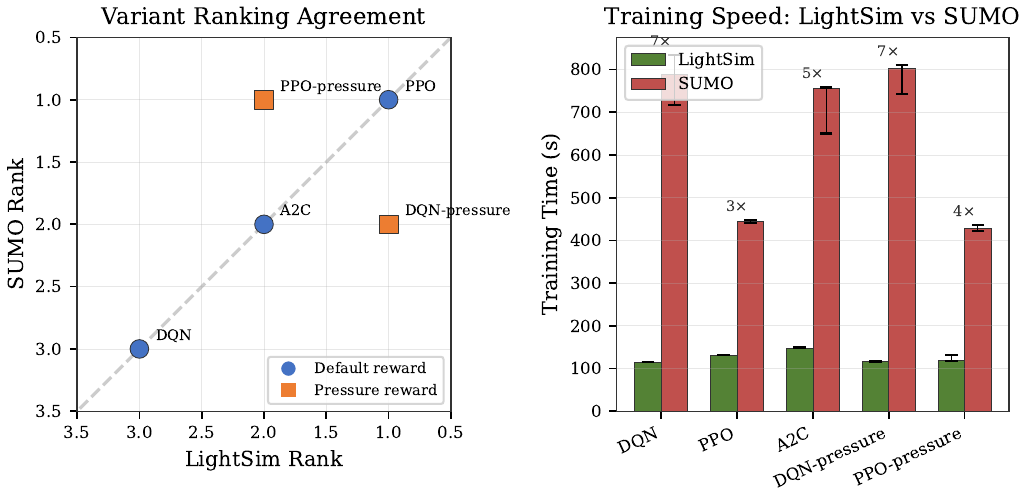}
    \caption{RL cross-validation: 5 variants $\times$ 5 seeds $\times$ 2 simulators. \textbf{Left:} Rank agreement---both simulators agree on PPO $>$ A2C $>$ DQN under the default reward. \textbf{Right:} Training time (median $\pm$ IQR); LightSim is $3$--$7\times$ faster.}
    \label{fig:rl_crossval}
\end{figure}

\paragraph{Ranking agreement.}
Figure~\ref{fig:rl_crossval} (left) shows the rank comparison.
Under the default reward, both simulators produce the \emph{identical} ranking: PPO $>$ A2C $>$ DQN.
Under the pressure reward, the two variants swap: LightSim ranks DQN-pressure above PPO-pressure, while SUMO ranks PPO-pressure above DQN-pressure.
This single disagreement occurs where the two variants' mean rewards are close relative to their cross-seed variance, making the ranking sensitive to stochastic factors.
Overall, 3 of 4 pairwise rankings agree across simulators, confirming that LightSim reliably identifies the best-performing RL algorithms for downstream evaluation in higher-fidelity environments.

\paragraph{Training speed.}
Figure~\ref{fig:rl_crossval} (right) compares median training times (with interquartile ranges across 5 seeds).
LightSim trains $3$--$7\times$ faster than SUMO across all variants: DQN trains in 116s vs.\ 789s ($6.8\times$), PPO in 132s vs.\ 444s ($3.4\times$), and A2C in 148s vs.\ 756s ($5.1\times$).
This speedup enables rapid hyperparameter sweeps and algorithm comparison that would be prohibitively slow in SUMO.

\paragraph{Caveats.}
A2C exhibits high instability in LightSim: 2 of 5 seeds fail to converge (reward $\approx -640{,}000$ vs.\ $\approx -44{,}000$ for converged seeds), compared to stable convergence across all 5 seeds in SUMO.
This is attributable to LightSim's deterministic dynamics, which create sharper optimization landscapes where A2C's high-variance gradient estimates can diverge more readily.
Additionally, reward scales differ substantially between simulators (LightSim operates on raw cumulative queue counts; SUMO on normalized per-step values), so only \emph{relative} rankings---not absolute reward magnitudes---should be compared.

\begin{figure*}[t]
    \centering
    \includegraphics[width=\linewidth]{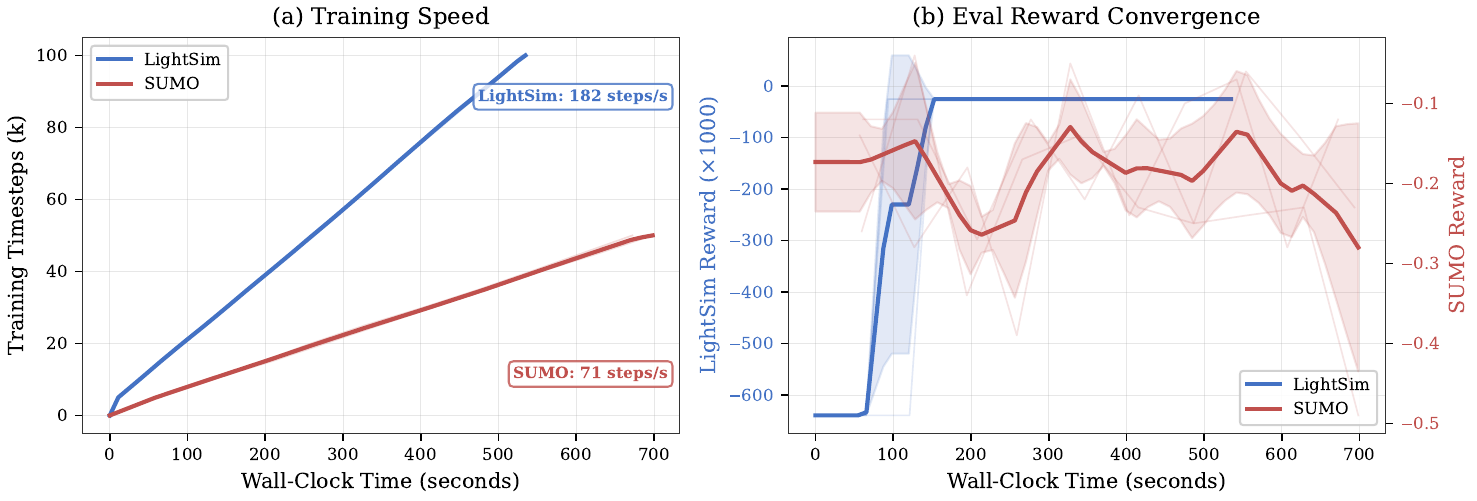}
    \caption{PPO training efficiency on single intersection (3 seeds, shaded $\pm 1$ std). (a)~LightSim processes 182 RL timesteps/s vs.\ SUMO's 71. (b)~Both simulators converge to near-optimal reward; LightSim reaches convergence within ${\sim}100$s.}
    \label{fig:sample_efficiency}
\end{figure*}

\paragraph{Sample efficiency.}
\label{sec:sample_efficiency}
This speed advantage translates directly into faster time-to-solution.
PPO is trained on the single-intersection scenario in both LightSim and SUMO with periodic evaluation every 5{,}000 timesteps (3 seeds; 100k steps for LightSim, 50k for SUMO).
Figure~\ref{fig:sample_efficiency} shows that LightSim processes RL training timesteps $2.6\times$ faster end-to-end (182 vs.\ 71 steps/s including evaluation checkpoints).
The raw simulation throughput advantage is even larger (${\sim}50\times$, see Table~\ref{tab:speed}); the gap narrows during RL training because neural network updates and evaluation episodes are simulator-independent overhead.
This speedup compounds over typical RL workflows: a hyperparameter sweep over 20 configurations $\times$ 5 seeds saves approximately 17 hours compared to SUMO, making LightSim practical for rapid algorithm development on a single machine.

\subsection{Sim-to-Sim Transfer}
\label{sec:transfer}

The preceding sections establish ranking agreement; a natural follow-up is whether policies learned in LightSim produce useful strategies in higher-fidelity environments.
A DQN agent is trained in LightSim on the single-intersection scenario (100k timesteps), its learned phase timing pattern is recorded, and the same timing is replayed in SUMO.
The RL agent learns an asymmetric green split of approximately 20s/17.5s (vs.\ the baseline's symmetric 30s/30s), reflecting the asymmetric demand (NS~$>$~EW).
Table~\ref{tab:transfer} shows that this LightSim-learned timing, when applied in SUMO, reduces average delay by $4.9\times$ (27.5s vs.\ 135.9s) and queue length by $2.1\times$ (5.5 vs.\ 11.8 vehicles) compared to the default fixed-time controller, while maintaining identical throughput.
Signal timing strategies discovered through rapid prototyping in LightSim can thus transfer effectively to microscopic simulation.

\begin{table}[t]
\caption{Sim-to-sim transfer: DQN timing learned in LightSim, evaluated in SUMO (single intersection, 3{,}600~s).}
\label{tab:transfer}
\centering
\small
\begin{tabular}{lrrr}
\toprule
\textbf{Controller (in SUMO)} & \textbf{Throughput} & \textbf{Avg.\ Delay (s)} & \textbf{Queue (veh)} \\
\midrule
FixedTime (30s/30s)                  & 3{,}522 & 135.9 & 11.8 \\
LightSim-learned DQN (20s/17.5s)    & 3{,}539 & 27.5  & 5.5  \\
\bottomrule
\end{tabular}
\end{table}

\subsection{Multi-Agent Evaluation}
\label{sec:multi_agent}

The preceding experiments focus on single-intersection control; scaling to network-level coordination is the next challenge.
LightSim's PettingZoo interface enables multi-agent evaluation on grid networks where each intersection is controlled by an independent agent.
The PettingZoo environment naturally handles \emph{heterogeneous} observation spaces---corner, edge, and center intersections have 14, 12, and 10 observation dimensions respectively, reflecting their different numbers of incoming links (corners have additional boundary demand links).

A shared-parameter DQN policy is trained on the grid-$4 \times 4$ (16 agents) using zero-padded observations and the independent learners paradigm.
A single policy network is trained by cycling through all agents' observations, then deployed identically at each intersection during evaluation.
Table~\ref{tab:multi_agent} shows results on 3{,}600-step episodes (5 episodes, compared against FixedTime and MaxPressure baselines).
The DQN agent achieves a per-step reward of $-233$, substantially outperforming FixedTime ($-1{,}238$) and MaxPressure ($-1{,}410$), while achieving $4.8\times$ higher throughput (15{,}528 vs.\ 3{,}229 vehicles).
Even simple independent learners with parameter sharing can thus learn effective multi-intersection coordination in LightSim.

\begin{table}[t]
\caption{Multi-agent DQN on grid-$4 \times 4$ (16 agents, 3{,}600 steps, shared policy with zero-padded observations).}
\label{tab:multi_agent}
\centering
\small
\begin{tabular}{lrrr}
\toprule
\textbf{Controller} & \textbf{Reward/step} & \textbf{Throughput} & \textbf{Queue} \\
\midrule
DQN (shared) & $-233$ & 15{,}528 & --- \\
FixedTime   & $-1{,}238$ & 3{,}229 & 2{,}452 \\
MaxPressure & $-1{,}410$ & 2{,}770 & 3{,}069 \\
\bottomrule
\end{tabular}
\end{table}

\paragraph{Multi-agent Decision Transformer.}
The evaluation extends to offline RL with a Decision Transformer (DT) \citep{chen2021decision} using parameter sharing on grid-$4 \times 4$.
A single DT model (64-dim, 3 layers, $\sim$418K parameters) is shared across all 16 agents, each maintaining its own rolling context buffer.
Observations are zero-padded to the maximum dimension (14).

Per-agent trajectories are collected from MaxPressure and GreenWave (the two best-performing classical controllers) with 40 episodes each, yielding 1{,}280 expert trajectories ($\sim$920K steps) across all 16 agents.
GreenWave offsets are computed row-wise from node coordinates to provide coordination signal in the training data.
The DT is trained on GPU for 10 epochs with behavioral cloning on the expert demonstrations.

Table~\ref{tab:multi_agent_dt} shows the results.
DT achieves a per-step reward of $-132$, outperforming all five rule-based baselines---including GreenWave ($-179$), the best classical controller---by 26\%, while achieving 8\% higher throughput (19{,}723 vs.\ 18{,}218 vehicles).
Notably, the DT outperforms both of its constituent expert controllers individually: MaxPressure ($-243$) and GreenWave ($-179$), suggesting that the model synthesizes complementary strengths from both strategies.

\begin{table}[t]
\caption{Multi-agent Decision Transformer on grid-$4 \times 4$ (16 agents, 720 steps, 10 episodes). Trained via behavioral cloning on MaxPressure + GreenWave demonstrations.}
\label{tab:multi_agent_dt}
\centering
\small
\begin{tabular}{lrrr}
\toprule
\textbf{Controller} & \textbf{Reward/step} & \textbf{Throughput} & \textbf{Vehicles} \\
\midrule
DT (expert)       & $\mathbf{-132}$  & \textbf{19{,}723} & 3{,}858 \\
\midrule
GreenWave         & $-179$  & 18{,}218 & 4{,}517 \\
FixedTime         & $-182$  & 18{,}056 & 4{,}540 \\
Webster           & $-197$  & 16{,}799 & 4{,}695 \\
SOTL              & $-220$  & 15{,}833 & 4{,}920 \\
MaxPressure       & $-243$  & 17{,}023 & 5{,}381 \\
\bottomrule
\end{tabular}
\end{table}

\paragraph{Multi-agent cross-validation.}
\label{sec:marl_crossval}
To test whether RL rankings extend to the multi-agent setting, DQN and PPO with shared parameters are trained on grid-$4 \times 4$ in both LightSim and SUMO (50{,}000 timesteps, 3 seeds, independent learners with observation padding).
LightSim ranks PPO above DQN (mean reward $-100{,}778$ vs.\ $-157{,}155$), while SUMO ranks DQN above PPO ($-645.7$ vs.\ $-743.7$), yielding Kendall's $\tau = -1.0$.
This reversal is expected: unlike the single-intersection case, the multi-agent wrappers necessarily differ between simulators---LightSim and SUMO use different observation spaces (density-based vs.\ queue-based), reward structures, and action semantics for multi-agent coordination.
With only two algorithms, a single swap produces the most extreme possible $\tau$ value.
The speed advantage remains substantial: LightSim trains PPO in 94s vs.\ SUMO's 906s ($9.6\times$ speedup).
This result highlights that while ranking preservation holds robustly for single-intersection and classical controllers, multi-agent RL cross-validation requires careful environment alignment to ensure comparable observation and reward definitions.

\subsection{Mesoscopic Validation}
\label{sec:mesoscopic_val}

The mesoscopic extensions introduced in Section~\ref{sec:mesoscopic} are designed to close the fidelity gap with SUMO.
Controller rankings are compared across three configurations: (i) LightSim default (deterministic, $\tau_L = 0$), (ii) LightSim mesoscopic (stochastic, $\tau_L = 2$s), and (iii) SUMO.

\begin{figure}[t]
    \centering
    \begin{subfigure}[b]{0.48\textwidth}
        \includegraphics[width=\textwidth]{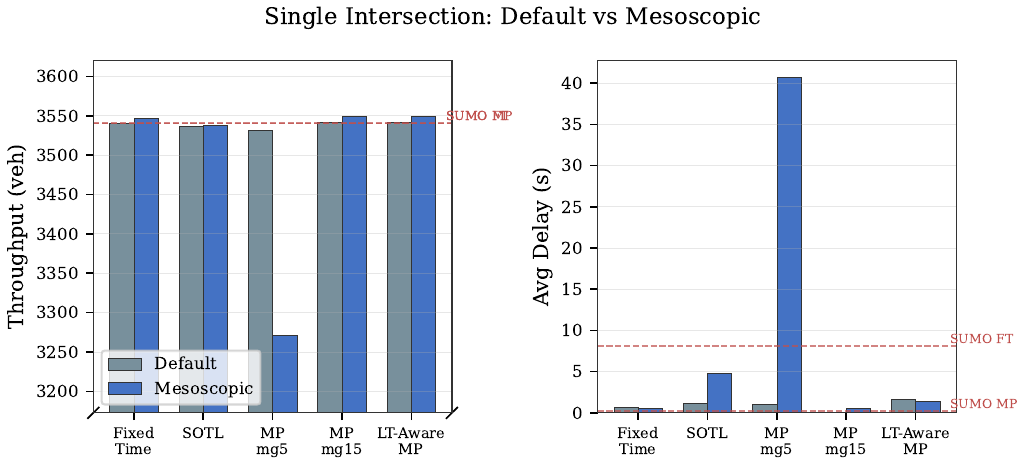}
        \caption{Single intersection}
        \label{fig:meso_crossval_single}
    \end{subfigure}
    \hfill
    \begin{subfigure}[b]{0.48\textwidth}
        \includegraphics[width=\textwidth]{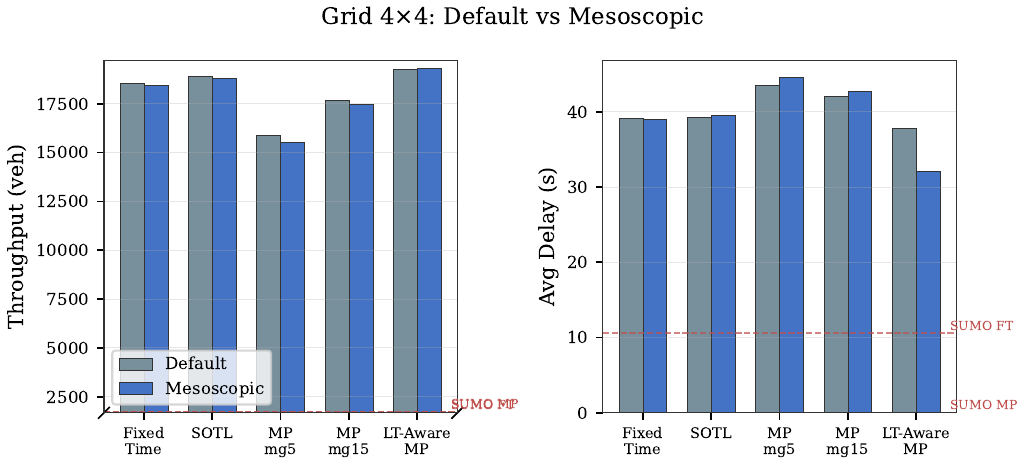}
        \caption{Grid $4 \times 4$}
        \label{fig:meso_crossval_grid}
    \end{subfigure}
    \caption{Controller throughput and delay across default vs.\ mesoscopic modes. Dashed red lines show SUMO reference values. \textbf{(a)} Mesoscopic mode reveals MaxPressure-mg5's vulnerability to switching cost. \textbf{(b)} LT-Aware MaxPressure achieves the highest throughput and lowest delay in mesoscopic mode.}
    \label{fig:meso_crossval}
\end{figure}

\paragraph{Controller ranking on single intersection.}
Table~\ref{tab:meso_single} compares five representative controllers on the single-intersection scenario (3{,}600 steps, 5 seeds for mesoscopic).
In the default (deterministic) mode, all controllers achieve similar throughput ($\sim$3{,}540 vehicles) and MaxPressure with $\text{mg}=5$ performs comparably to others.
When mesoscopic extensions are enabled, MaxPressure-mg5 collapses: delay jumps from 1.1s to 40.7s and throughput drops by 8\%.
With 5s minimum green and 7s dead time per switch (3s yellow $+$ 2s all-red $+$ 2s lost time), MaxPressure-mg5 achieves only 42\% effective green ratio.
In contrast, LT-Aware MaxPressure avoids unnecessary switches and maintains low delay (1.4s), matching SUMO's actuated MaxPressure ranking.
Figure~\ref{fig:meso_crossval} visualizes these results.

\begin{table}[t]
\caption{Mesoscopic cross-validation on single intersection (3{,}600~s). MaxPressure-mg5 collapses under lost time ($\tau_L = 2$s); LT-Aware MP avoids this by incorporating switching cost.}
\label{tab:meso_single}
\centering
\small
\begin{tabular}{llrrr}
\toprule
\textbf{Mode} & \textbf{Controller} & \textbf{Throughput} & \textbf{Delay (s)} & \textbf{Queue} \\
\midrule
\multirow{5}{*}{Default}
& FixedTime-30s      & 3{,}540 & 0.67  & 17.1 \\
& SOTL               & 3{,}537 & 1.13  & 18.9 \\
& MaxPressure-mg5    & 3{,}531 & 1.09  & 18.8 \\
& MaxPressure-mg15   & 3{,}542 & 0.00  & 0.0  \\
& LT-Aware-MP        & 3{,}541 & 1.64  & 21.0 \\
\midrule
\multirow{5}{*}{Mesoscopic}
& FixedTime-30s      & 3{,}547 & 0.61  & 9.6  \\
& SOTL               & 3{,}538 & 4.77  & 32.0 \\
& MaxPressure-mg5    & 3{,}271 & 40.73 & 227.7 \\
& MaxPressure-mg15   & 3{,}549 & 0.53  & 9.3  \\
& LT-Aware-MP        & 3{,}549 & 1.42  & 14.3 \\
\midrule
\multirow{2}{*}{SUMO}
& FixedTime          & 3{,}540 & 8.13  & 28.0 \\
& MaxPressure        & 3{,}541 & 0.19  & 4.0  \\
\bottomrule
\end{tabular}
\end{table}

\paragraph{Grid $4 \times 4$ validation.}
On the 16-intersection grid, LT-Aware MaxPressure dominates in both modes, achieving 19{,}329 vehicles exited and 32.1s average delay under mesoscopic conditions---the only controller to \emph{improve} throughput when lost time is introduced, by eliminating wasteful phase switches.
Standard MaxPressure-mg5 is the worst performer in both modes (15{,}539 exited, 44.6s delay in mesoscopic).
This ranking is consistent with SUMO, where actuated MaxPressure (analogous to our LT-Aware variant) outperforms FixedTime (Figure~\ref{fig:meso_crossval_grid}).

\begin{figure}[t]
    \centering
    \begin{subfigure}[b]{0.48\textwidth}
        \includegraphics[width=\textwidth]{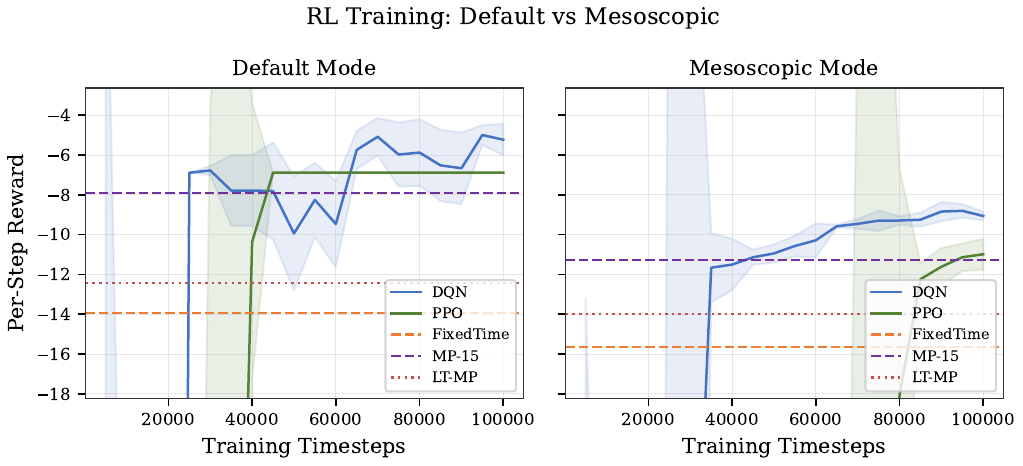}
        \caption{Learning curves}
        \label{fig:meso_rl_curves}
    \end{subfigure}
    \hfill
    \begin{subfigure}[b]{0.48\textwidth}
        \includegraphics[width=\textwidth]{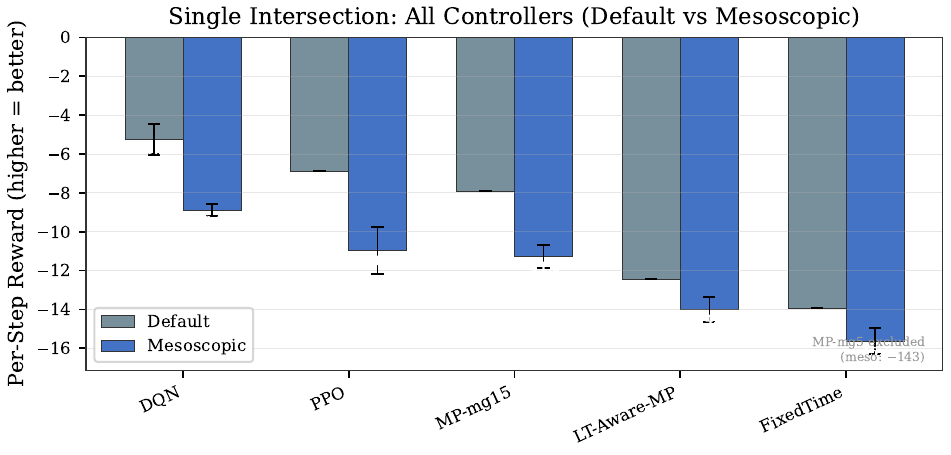}
        \caption{Final performance}
        \label{fig:meso_summary}
    \end{subfigure}
    \caption{RL training under default vs.\ mesoscopic mode. \textbf{(a)} DQN and PPO converge within 60k timesteps in both modes. \textbf{(b)} DQN is the best controller in both modes; mesoscopic mode degrades all methods but preserves the ranking.}
    \label{fig:meso_rl}
\end{figure}

\paragraph{RL training under mesoscopic conditions.}
DQN and PPO agents (100k timesteps, 5 seeds each) are trained in both default and mesoscopic modes on the single intersection.
Table~\ref{tab:meso_rl} and Figure~\ref{fig:meso_rl} show the results.
In default mode, DQN achieves $-5.23 \pm 0.79$ per-step reward, outperforming all baselines.
In mesoscopic mode, DQN still outperforms all rule-based controllers ($-8.89 \pm 0.32$), though the gap narrows as stochastic demand and lost time increase the difficulty.
PPO shows higher variance in mesoscopic mode ($\pm 1.21$ vs.\ $\pm 0.00$), reflecting the additional stochasticity.
Both RL agents consistently outperform all baselines in both modes.

\begin{table}[t]
\caption{RL and baseline performance under default vs.\ mesoscopic modes (per-step reward $\pm$ std; higher = better). Mesoscopic mode increases difficulty but preserves the ranking.}
\label{tab:meso_rl}
\centering
\small
\begin{tabular}{llrr}
\toprule
\textbf{Mode} & \textbf{Controller} & \textbf{Reward} & \textbf{Throughput} \\
\midrule
\multirow{4}{*}{Default}
& DQN            & $-5.23 \pm 0.79$  & 3{,}542 \\
& PPO            & $-6.89 \pm 0.00$  & 3{,}542 \\
& MaxPressure-15 & $-7.90$            & 3{,}542 \\
& FixedTime      & $-13.94$           & 3{,}540 \\
\midrule
\multirow{4}{*}{Mesoscopic}
& DQN            & $-8.89 \pm 0.32$  & 3{,}540 \\
& PPO            & $-10.96 \pm 1.21$ & 3{,}540 \\
& MaxPressure-15 & $-11.27$           & 3{,}557 \\
& FixedTime      & $-15.64$           & 3{,}556 \\
\bottomrule
\end{tabular}
\end{table}

\subsection{Real-World Network Evaluation}
\label{sec:osm_eval}

To verify that controller rankings generalize beyond synthetic topologies, five controllers are evaluated across six of the sixteen built-in OpenStreetMap city networks, selected to span four continents and a range of network sizes: Manhattan (44 signals), Shanghai (73 signals), London (131 signals), San Francisco (55 signals), Mumbai (40 signals), and Sioux Falls (43 signals).
Each combination is run for 3{,}600 steps with 3 stochastic demand seeds; Table~\ref{tab:osm_cities} reports mean throughput $\pm$ standard deviation.

\begin{table}[t]
\caption{Controller throughput (vehicles exited, mean $\pm$ std over 3 seeds) on six OSM city networks. MaxPressure ranks first in 4 of 6 cities.}
\label{tab:osm_cities}
\centering
\small
\begin{tabular}{lrrrrr}
\toprule
\textbf{City} & \textbf{FixedTime} & \textbf{Webster} & \textbf{MaxPressure} & \textbf{SOTL} & \textbf{GreenWave} \\
\midrule
Manhattan & 11{,}783\tiny{$\pm$111} & 11{,}657\tiny{$\pm$128} & \textbf{11{,}956}\tiny{$\pm$121} & 11{,}842\tiny{$\pm$121} & 11{,}743\tiny{$\pm$103} \\
Shanghai & 22{,}778\tiny{$\pm$153} & 22{,}432\tiny{$\pm$186} & 22{,}653\tiny{$\pm$165} & 22{,}773\tiny{$\pm$160} & \textbf{22{,}784}\tiny{$\pm$147} \\
London & 44{,}421\tiny{$\pm$19} & 43{,}845\tiny{$\pm$42} & \textbf{45{,}355}\tiny{$\pm$60} & 43{,}697\tiny{$\pm$15} & 44{,}611\tiny{$\pm$25} \\
San Francisco & 20{,}011\tiny{$\pm$116} & 19{,}957\tiny{$\pm$97} & 20{,}118\tiny{$\pm$129} & \textbf{20{,}348}\tiny{$\pm$143} & 19{,}998\tiny{$\pm$118} \\
Mumbai & 15{,}356\tiny{$\pm$141} & 15{,}464\tiny{$\pm$105} & \textbf{15{,}523}\tiny{$\pm$117} & 15{,}369\tiny{$\pm$123} & 15{,}208\tiny{$\pm$130} \\
Sioux Falls & 15{,}363\tiny{$\pm$121} & 15{,}293\tiny{$\pm$127} & \textbf{15{,}500}\tiny{$\pm$137} & 14{,}926\tiny{$\pm$233} & 15{,}185\tiny{$\pm$103} \\
\bottomrule
\end{tabular}
\end{table}

MaxPressure ranks first in 4 of 6 cities (Manhattan, London, Mumbai, Sioux Falls), while GreenWave leads in Shanghai and SOTL in San Francisco---confirming that adaptive controllers are competitive across diverse topologies, though no single strategy dominates universally.
The mean pairwise Kendall's $\tau$ across all $\binom{6}{2}=15$ city pairs is $\bar{\tau}=0.12$ for the full 5-controller ranking, reflecting topology-dependent variation.
Nevertheless, MaxPressure is the most consistent top performer (first in 4 of 6 cities), while Webster ranks last in three of six cities, as its pre-computed cycle lengths are less effective on irregular real-world topologies.

\section{Discussion and Limitations}

\paragraph{When to use LightSim.}
LightSim is best suited as a \emph{fast prototyping environment that preserves algorithmic rankings}: researchers can rapidly iterate on RL algorithms and reward designs in LightSim, confident that the relative performance ordering will transfer to higher-fidelity simulators like SUMO.
Its macroscopic dynamics faithfully capture queue formation and discharge, which are the primary phenomena that signal control algorithms must learn to manage.
The mesoscopic extensions add realistic phase-switching penalties and demand stochasticity, making LightSim suitable for evaluating controllers that are sensitive to these factors.
For researchers developing new RL algorithms for signal control, LightSim offers $3$--$7\times$ faster training than SUMO with ranking-preserving fidelity, as confirmed across single-intersection, arterial, and multi-agent grid topologies (Sections~\ref{sec:rl_crossval}--\ref{sec:marl_crossval}).

\paragraph{When not to use LightSim.}
LightSim does not model individual vehicle trajectories, lane-changing, or detailed intersection geometry.
Research requiring vehicle-level metrics (e.g., fuel consumption models, safety analysis with time-to-collision), mixed autonomy (human and autonomous vehicles), or detailed geometric intersection design should use microscopic simulators.

\paragraph{Limitations.}
\begin{itemize}
    \item \textbf{Macroscopic fidelity.} The CTM assumes homogeneous traffic and symmetric fundamental diagrams. Real traffic exhibits heterogeneous vehicle types and asymmetric capacity drops. The mesoscopic extensions address the most impactful omissions (lost time, stochastic arrivals) but do not model lane-changing, platoon dispersion, or turning movement conflicts.
    \item \textbf{Network scale and speed crossover.} LightSim is $4$--$6\times$ faster than SUMO for small to medium networks (1--20 intersections), but the pure-Python implementation incurs overhead that narrows this gap at larger scales. On the $8 \times 8$ grid (64 intersections), SUMO's compiled C++ engine matches LightSim's speed (Table~\ref{tab:sumo}). For city-wide networks with thousands of intersections, a compiled backend (e.g., Cython, JAX, or Rust) would be required. We emphasize that LightSim's primary advantage for large networks is \emph{simplicity} (no IPC, no XML, pip-installable), not raw throughput.
    \item \textbf{Transfer gap.} Policies trained in LightSim operate on aggregate densities rather than individual vehicles. While the learned signal timing strategies can transfer to SUMO (Section~\ref{sec:transfer}), fine-tuning in a higher-fidelity simulator may be needed for real-world deployment. We view LightSim as a rapid prototyping tool, not a replacement for SUMO or real-world testing.
    \item \textbf{Absolute metric divergence.} While controller \emph{rankings} are consistent between LightSim and SUMO, absolute delay and queue values diverge substantially---often by an order of magnitude---due to the fundamentally different modeling approaches (aggregate density vs.\ individual vehicle tracking). This divergence is inherent to any macroscopic-vs-microscopic comparison and means that LightSim cannot be used for absolute delay calibration. Users should compare relative performance, not absolute numbers.
    \item \textbf{RL ranking edge cases.} Our RL cross-validation shows perfect ranking agreement under the default reward (3/3 pairs), but the pressure reward variants swap ranks between simulators (Section~\ref{sec:rl_crossval}). Rankings may disagree when algorithms produce similar mean rewards and the difference falls within cross-seed variance.
\end{itemize}

\paragraph{Future work.}
Natural extensions include GPU-accelerated CTM computation via JAX or PyTorch for batched environment simulation; origin--destination matrix demand models for more realistic trip generation; multi-modal extensions (buses, pedestrians); and integration with sim-to-real transfer frameworks for bridging the gap from LightSim to microscopic simulators and real-world deployments.

\section{Conclusion}

We introduced LightSim, a lightweight, pip-installable traffic signal simulator built on the Cell Transmission Model.
LightSim provides standard Gymnasium and PettingZoo interfaces, achieves thousands of simulation steps per second, and reproduces the theoretical fundamental diagram exactly.
Mesoscopic extensions---start-up lost time and stochastic demand---close the fidelity gap with microscopic simulators, while RL agents (DQN, PPO) consistently outperform all rule-based baselines.

Systematic cross-simulator experiments---spanning single intersections, arterial corridors, grid networks, and six real-world OpenStreetMap cities---confirm that LightSim preserves controller rankings from SUMO for both classical and RL strategies while training $3$--$7\times$ faster.
These results validate LightSim's role as a fast prototyping environment that faithfully identifies the best-performing algorithms for downstream evaluation in higher-fidelity simulators.

LightSim is released as an open-source benchmark with nineteen built-in scenarios across four continents, seven baseline controllers, a web visualization dashboard, and full RL training pipelines, with the goal of lowering the barrier to traffic signal control research from days to minutes.

\begin{ack}
The authors thank the open-source communities behind NumPy, Gymnasium, PettingZoo, Stable-Baselines3, and SUMO for providing the foundations on which LightSim is built.
\end{ack}

\bibliographystyle{unsrtnat}
\bibliography{references}

\newpage
\appendix
\section*{Appendix}

\subsection*{A. Broader Impact}

LightSim is a research tool for developing traffic signal control algorithms.
Improved signal timing has direct positive societal impacts: reduced congestion, lower emissions, and shorter travel times.
We do not foresee negative societal impacts from this work, beyond the general concern that simulation-trained policies require careful validation before real-world deployment.

\subsection*{B. Code and Data Availability}

LightSim is open-source under the MIT license. Code, pretrained weights, and all experiment scripts are available at \url{https://github.com/AnthonySu/LightSim}.

\subsection*{C. Reproducibility}

All experiments can be reproduced using the scripts provided in the repository:
\begin{verbatim}
    pip install lightsim[all]
    python -m lightsim.benchmarks.speed_benchmark
    python -m lightsim.benchmarks.rl_baselines --train-rl --timesteps 100000
    python -m lightsim.benchmarks.sumo_comparison
    python scripts/cross_validation_mesoscopic.py
    python scripts/rl_mesoscopic_experiment.py
    python scripts/rl_cross_validation.py
    python scripts/generate_figures.py
\end{verbatim}

A complete mapping of scripts to paper figures and tables is provided in \texttt{scripts/README.md}.
LightSim's simulation is deterministic given a random seed, ensuring exact reproducibility.
The mesoscopic mode uses seeded NumPy random generators for reproducible stochastic runs.
Pretrained RL checkpoints (DQN and PPO on single-intersection, with both queue and pressure rewards) are included in the repository under \texttt{weights/}, enabling immediate evaluation without retraining.

\end{document}

%% file: tables/ft_reward.tex
$-13.94$

%% file: tables/ft_throughput.tex
3{,}540

%% file: tables/ft_delay.tex
0.67

%% file: tables/mp_reward.tex
$-24.60$

%% file: tables/mp_throughput.tex
3{,}531

%% file: tables/mp_delay.tex
1.09

%% file: tables/dqn_reward.tex
$-5.23 \pm 0.79$

%% file: tables/dqn_throughput.tex
3{,}542

%% file: tables/dqn_delay.tex
$0.08 \pm 0.10$

%% file: tables/ppo_reward.tex
$-6.89 \pm 0.00$

%% file: tables/ppo_throughput.tex
3{,}542

%% file: tables/ppo_delay.tex
$0.00 \pm 0.00$

%% file: references.bib
@article{wei2018intellilight,
  title={{IntelliLight}: A reinforcement learning approach for intelligent traffic light control},
  author={Wei, Hua and Zheng, Guanjie and Yao, Huaxiu and Li, Zhenhui},
  journal={Proceedings of the 24th ACM SIGKDD International Conference on Knowledge Discovery \& Data Mining},
  pages={2496--2505},
  year={2018},
  publisher={ACM}
}

@inproceedings{wei2019presslight,
  title={{PressLight}: Learning max pressure control to coordinate traffic signals in arterial network},
  author={Wei, Hua and Chen, Chacha and Zheng, Guanjie and Wu, Kan and Gayah, Vikash and Xu, Kai and Li, Zhenhui},
  booktitle={Proceedings of the 25th ACM SIGKDD International Conference on Knowledge Discovery \& Data Mining},
  pages={1290--1298},
  year={2019}
}

@article{wei2021recent,
  title={Recent advances in reinforcement learning for traffic signal control: A survey of models and evaluation},
  author={Wei, Hua and Zheng, Guanjie and Gayah, Vikash and Li, Zhenhui},
  journal={ACM SIGKDD Explorations Newsletter},
  volume={22},
  number={2},
  pages={12--18},
  year={2021}
}

@inproceedings{zheng2019learning,
  title={Learning phase competition for traffic signal control},
  author={Zheng, Guanjie and Xiong, Yuanhao and Zang, Xinshi and Feng, Jie and Wei, Hua and Zhang, Huichu and Li, Yong and Xu, Kai and Li, Zhenhui},
  booktitle={Proceedings of the 28th ACM International Conference on Information and Knowledge Management},
  pages={1963--1972},
  year={2019}
}

@article{chen2020toward,
  title={Toward a thousand lights: Decentralized deep reinforcement learning for large-scale traffic signal control},
  author={Chen, Chacha and Wei, Hua and Xu, Nan and Zheng, Guanjie and Yang, Ming and Xiong, Yuanhao and Xu, Kai and Li, Zhenhui},
  journal={Proceedings of the AAAI Conference on Artificial Intelligence},
  volume={34},
  number={04},
  pages={3414--3421},
  year={2020}
}

@article{lopez2018microscopic,
  title={Microscopic traffic simulation using {SUMO}},
  author={Lopez, Pablo Alvarez and Behrisch, Michael and Bieker-Walz, Laura and Erdmann, Jakob and Fl{\"o}tter{\"o}d, Yun-Pang and Hilbrich, Robert and L{\"u}cken, Leonhard and Rummel, Johannes and Wagner, Peter and Wie{\ss}ner, Evamarie},
  journal={Proceedings of the 21st IEEE International Conference on Intelligent Transportation Systems},
  pages={2575--2582},
  year={2018},
  publisher={IEEE}
}

@article{zhang2019cityflow,
  title={{CityFlow}: A multi-agent reinforcement learning environment for large scale city traffic signal control},
  author={Zhang, Huichu and Feng, Siyuan and Liu, Chang and Ding, Yaoyao and Zhu, Yichen and Zhou, Zihan and Zhang, Weinan and Yu, Yong and Jin, Haiming and Li, Zhenhui},
  journal={Proceedings of the World Wide Web Conference},
  pages={3620--3624},
  year={2019}
}

@inproceedings{vinitsky2018benchmarks,
  title={Benchmarks for reinforcement learning in mixed-autonomy traffic},
  author={Vinitsky, Eugene and Kreidieh, Aboudy and Le Flem, Luc and Kheterpal, Nishant and Jang, Kanaad and Wu, Cathy and Wu, Francis and Liaw, Richard and Liang, Eric and Bayen, Alexandre M},
  booktitle={Conference on Robot Learning},
  pages={399--409},
  year={2018},
  organization={PMLR}
}

@article{wu2021flow,
  title={Flow: A modular learning framework for mixed autonomy traffic},
  author={Wu, Cathy and Kreidieh, Aboudy R and Parvate, Kanaad and Vinitsky, Eugene and Bayen, Alexandre M},
  journal={IEEE Transactions on Robotics},
  volume={38},
  number={2},
  pages={1270--1286},
  year={2021}
}

@article{daganzo1994cell,
  title={The cell transmission model: A dynamic representation of highway traffic consistent with the hydrodynamic theory},
  author={Daganzo, Carlos F},
  journal={Transportation Research Part B: Methodological},
  volume={28},
  number={4},
  pages={269--287},
  year={1994},
  publisher={Elsevier}
}

@article{daganzo1995cell,
  title={The cell transmission model, part {II}: Network traffic},
  author={Daganzo, Carlos F},
  journal={Transportation Research Part B: Methodological},
  volume={29},
  number={2},
  pages={79--93},
  year={1995},
  publisher={Elsevier}
}

@article{lighthill1955kinematic,
  title={On kinematic waves {II}. A theory of traffic flow on long crowded roads},
  author={Lighthill, Michael James and Whitham, Gerald Beresford},
  journal={Proceedings of the Royal Society of London. Series A. Mathematical and Physical Sciences},
  volume={229},
  number={1178},
  pages={317--345},
  year={1955},
  publisher={The Royal Society London}
}

@article{richards1956shock,
  title={Shock waves on the highway},
  author={Richards, Paul I},
  journal={Operations Research},
  volume={4},
  number={1},
  pages={42--51},
  year={1956},
  publisher={INFORMS}
}

@article{varaiya2013max,
  title={Max pressure control of a network of signalized intersections},
  author={Varaiya, Pravin},
  journal={Transportation Research Part C: Emerging Technologies},
  volume={36},
  pages={177--195},
  year={2013},
  publisher={Elsevier}
}

@article{raffin2021stable,
  title={Stable-{Baselines3}: Reliable reinforcement learning implementations},
  author={Raffin, Antonin and Hill, Ashley and Gleave, Adam and Kanervisto, Anssi and Ernestus, Maximilian and Dormann, Noah},
  journal={Journal of Machine Learning Research},
  volume={22},
  number={268},
  pages={1--8},
  year={2021}
}

@article{schulman2017proximal,
  title={Proximal policy optimization algorithms},
  author={Schulman, John and Wolski, Filip and Dhariwal, Prafulla and Radford, Alec and Klimov, Oleg},
  journal={arXiv preprint arXiv:1707.06347},
  year={2017}
}

@article{mnih2015human,
  title={Human-level control through deep reinforcement learning},
  author={Mnih, Volodymyr and Kavukcuoglu, Koray and Silver, David and Rusu, Andrei A and Veness, Joel and Bellemare, Marc G and Graves, Alex and Riedmiller, Martin and Fidjeland, Andreas K and Ostrovski, Georg and others},
  journal={Nature},
  volume={518},
  pages={529--533},
  year={2015},
  publisher={Nature Publishing Group}
}

@article{towers2024gymnasium,
  title={Gymnasium: A standard interface for reinforcement learning environments},
  author={Towers, Mark and Kwiatkowski, Ariel and Terry, Jordan and Balis, John U and De Cola, Gianluca and Deleu, Tristan and Goul{\~a}o, Manuel and Kallinteris, Andreas and KG, Arjun and Krimmel, Markus and others},
  journal={arXiv preprint arXiv:2407.17032},
  year={2024}
}

@article{terry2021pettingzoo,
  title={{PettingZoo}: Gym for multi-agent reinforcement learning},
  author={Terry, J and Black, Benjamin and Grammel, Nathaniel and Jayakumar, Mario and Hari, Ananth and Sullivan, Ryan and Santos, Luis S and Dieffendahl, Clemens and Horsch, Caroline and Perez-Vicente, Rodrigo and others},
  journal={Advances in Neural Information Processing Systems},
  volume={34},
  pages={15032--15043},
  year={2021}
}

@book{treiber2013traffic,
  title={Traffic Flow Dynamics: Data, Models and Simulation},
  author={Treiber, Martin and Kesting, Arne},
  year={2013},
  publisher={Springer-Verlag Berlin Heidelberg}
}

@inproceedings{oroojlooy2020attendlight,
  title={{AttendLight}: Universal attention-based reinforcement learning model for traffic signal control},
  author={Oroojlooy, Afshin and Nazari, Mohammadreza and Hajinezhad, Davood and Silva, Jorge},
  booktitle={Advances in Neural Information Processing Systems},
  volume={33},
  pages={4079--4090},
  year={2020}
}

@techreport{webster1958traffic,
  title={Traffic signal settings},
  author={Webster, FV},
  institution={Road Research Laboratory, UK},
  number={Technical Paper No. 39},
  year={1958}
}

@article{gershenson2005self,
  title={Self-organizing traffic lights},
  author={Gershenson, Carlos},
  journal={Complex Systems},
  volume={16},
  number={1},
  pages={29--53},
  year={2005}
}

@book{hcm2010,
  title={Highway Capacity Manual},
  author={{Transportation Research Board}},
  edition={5th},
  year={2010},
  publisher={National Academies Press},
  address={Washington, D.C.}
}

@inproceedings{wei2019colight,
  title={{CoLight}: Learning network-level cooperation for traffic signal control},
  author={Wei, Hua and Xu, Nan and Zhang, Huichu and Zheng, Guanjie and Zang, Xinshi and Chen, Chacha and Zhang, Weinan and Zhu, Yichen and Xu, Kai and Li, Zhenhui},
  booktitle={Proceedings of the 28th ACM International Conference on Information and Knowledge Management},
  pages={1913--1922},
  year={2019}
}

@inproceedings{chen2021decision,
  title={Decision Transformer: Reinforcement learning via sequence modeling},
  author={Chen, Lili and Lu, Kevin and Rajeswaran, Aravind and Lee, Kimin and Grover, Aditya and Laskin, Misha and Abbeel, Pieter and Srinivas, Aravind and Mordatch, Igor},
  booktitle={Advances in Neural Information Processing Systems},
  volume={34},
  pages={15084--15097},
  year={2021}
}

@inproceedings{su2022emvlight,
  title={{EMVLight}: A Decentralized Reinforcement Learning Framework for Efficient Passage of Emergency Vehicles},
  author={Su, Haoran and Zhong, Yaofeng D. and Chow, Joseph Y.J. and Dey, Biswadip and Jin, Li},
  booktitle={Proceedings of the AAAI Conference on Artificial Intelligence},
  volume={36},
  number={4},
  pages={4610--4618},
  year={2022}
}

@article{su2026dqjl,
  title={Hierarchical {GNN}-Based Multi-Agent Learning for Dynamic Queue-Jump Lane and Emergency Vehicle Corridor Formation},
  author={Su, Haoran},
  journal={arXiv preprint arXiv:2601.04177},
  year={2026}
}
